\documentclass[sigconf]{acmart}
\usepackage{graphicx}

\AtBeginDocument{%
  }

\copyrightyear{2025}
\acmYear{2025}
\setcopyright{acmlicensed}
\acmConference[WWW Companion '25] {Companion of the 16th ACM/SPEC International Conference on Performance Engineering}{April 28-May 2, 2025}{Sydney, NSW, Australia.}
\acmBooktitle{Companion of the 16th ACM/SPEC International Conference on Performance Engineering (WWW Companion '25), April 28-May 2, 2025, Sydney, NSW, Australia}
\acmISBN{979-8-4007-1331-6/25/04}
\acmDOI{10.1145/3701716.3715541}

\begin{document}

%% 去掉ACM Reference Format，camera ready需要打开
\settopmatter{printacmref=true}

\title{SQLord: A Robust Enterprise Text-to-SQL Solution via \\ Reverse Data Generation and Workflow Decomposition}

\author{Song Cheng}
\authornote{Corresponding author.}
\affiliation{%
  \institution{Alibaba Group}
  \city{Hangzhou, Zhejiang, China}
  \country{}}
\email{chengsong.cs@alibaba-inc.com}

\author{Qiannan Cheng}
\affiliation{%
  \institution{Alibaba Group}
  \city{Hangzhou, Zhejiang, China}
  \country{}}
\email{qiannan.cqn@alibaba-inc.com}

\author{Linbo Jin}
\affiliation{%
  \institution{Alibaba Group}
  \city{Hangzhou, Zhejiang, China}
  \country{}}
\email{yuyi.jlb@alibaba-inc.com}

\author{Lei Yi}
\affiliation{%
  \institution{Alibaba Group}
  \city{Hangzhou, Zhejiang, China}
  \country{}}
\email{yilei.yi@alibaba-inc.com}

\author{Guannan Zhang}
\affiliation{%
  \institution{Alibaba Group}
  \city{Hangzhou, Zhejiang, China}
  \country{}}
\email{zgn138592@alibaba-inc.com}

\renewcommand{\shortauthors} {cheng et al.}

\begin{abstract}
    Transforming natural language into SQL queries (NL2SQL) is crucial for data-driven business applications. Existing frameworks, trained on open-source datasets, struggle with complex business logic and lack domain-specific data for fine-tuning. Additionally, evaluation methods often require annotated data and executable database environments, which are scarce in real-world scenarios. To address these challenges, we propose \textit{SQLord}, an enterprise-level NL2SQL framework. First, SQLord introduces a data reverse generation approach to convert raw SQL statements into annotated data for supervised fine-tuning (SFT). Second, it proposes a decomposition method for complex queries using an automated workflow generator. Additionally, SQLord features a comprehensive GPT-Judge evaluation framework, including Execution Evaluation (EXE), Query-SQL Evaluation (QSE), and SQL-SQL Evaluation (SSE), tailored to diverse scenarios. Offline tests significantly outperform state-of-the-art baselines, and online accuracy consistently exceeds 90, highlighting SQLord’s advantages and effectiveness in complex real-world scenarios. SQLord has been successfully applied across multiple scenarios on the world's largest B2B e-commerce platform.
\end{abstract}

\begin{CCSXML}
  <ccs2012>
  <concept>
  <concept_id>10010147.10010178.10010179</concept_id>
  <concept_desc>Computing methodologies~Natural language processing</concept_desc>
  <concept_significance>500</concept_significance>
  </concept>
  </ccs2012>
\end{CCSXML}

\ccsdesc[500]{Computing methodologies~Natural language processing}

\keywords{text to sql, llm, workflow generation}

\maketitle

\begin{figure}[h]
\centering
\includegraphics[width=1.0\linewidth]{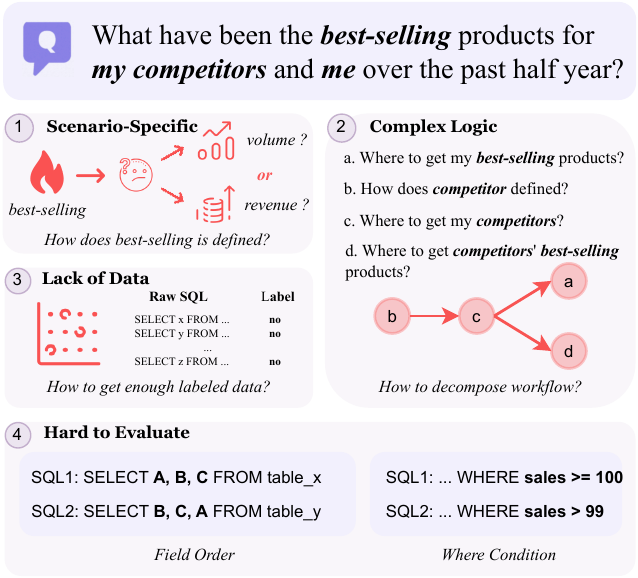}
\caption{A real-world online case of NL2SQL challenges}
\vspace{-1em}
\label{fig:introduce_pic}
\end{figure}

\section{Introduction}
Nature language to SQL (NL2SQL), which translates natural language inquiries into SQL statments is still a critical challange in both natural language processing and relational database communities \cite{DBLP:journals/corr/abs-2408-05109}. Historically, the researches in this domain \cite{DBLP:journals/corr/abs-2203-06958, DBLP:conf/aaai/Li00023, DBLP:journals/corr/abs-2301-07507, DBLP:conf/acl/WangSLPR20} has predominantly focused on identifying and abstracting patterns that correlate questions with SQL, subsequently refining these patterns through the deployment of encoder-decoder architectures trained on NL2SQL datasets. The emergence of large language models (LLMs) has marked a paradigm shift in methodologies for addressing the NL2SQL challenge \cite{DBLP:journals/corr/abs-2204-00498, DBLP:journals/pvldb/Trummer22, DBLP:conf/nips/PourrezaR23, gao2024text}. A notable contribution in this field was made by Gao et al. \cite{gao2024text}, who leveraged GPT-4 \cite{DBLP:journals/corr/abs-2303-08774} to achieve first place on the Spider leaderboard \cite{spider}, demonstrating an execution accuracy of 86.6\%. Diverging from traditional approaches, the principal challenge in LLM-based NL2SQL frameworks involves the strategic employment of prompt engineering to elicit precise SQL query generation. This process requires meticulous optimization of question representations \cite{DBLP:journals/corr/abs-2307-07306, DBLP:conf/nips/PourrezaR23}, careful selection of illustrative examples \cite{DBLP:conf/acl-deelio/LiuSZDCC22, nan2023enhancing}, and the systematic organization of them.

Despite advancements in LLM-based methods for natural language understanding and SQL query generation, real-world enterprise applications still face significant challenges. For example, as shown in Figure \ref{fig:introduce_pic}, a user of an online B2B e-commerce platform asking about the best-selling products of both competitors and themselves over the past six months presents several difficulties. First, defining "best-selling" (e.g., by sales volume or revenue) requires domain-specific knowledge for understanding business logic, such as identifying competitors and obtaining relevant data for the task. Second, LLMs' in-context learning capabilities are often inadequate for complex NL2SQL tasks. While fine-tuning the models could improve performance, however, there is a lack of labeled SQL queries for training, as most queries in real-world applications are not labeled. Third, evaluating NL2SQL performance is challenging due to variations in SQL queries, such as differences in field order or conditions (e.g., "sales >= 100" vs. "sales > 99"), which complicate direct string matching and performance assessment, especially when queries return empty results.

To address the challenges of applying NL2SQL in real-world scenarios, we propose \text{SQLord}. First, we introduce a data reverse generation approach that leverages SQL statements developed by developers, along with their comments, to generate annotated \texttt{<SQL, COMMENT>} pairs. These annotated pairs are then used to train the reverse generation model, \( RevLLM \), which generates natural language descriptions (referred to as "queries") for SQL statements. \( RevLLM \) facilitates the creation of a robust set of \texttt{<Query, SQL>} pairs from raw SQL commonly written in daily development, effectively mitigating the data scarcity issue for fine-tuning NL2SQL models in specialized domains. Second, we present a task decomposition strategy that combines domain-specific knowledge with automated workflow generation. This method systematically decomposes the original query into sub-tasks, enabling the handling of complex business logic, such as multi-table joins and SQL nesting, which are common in real-world applications. Third, we propose a flexible evaluation framework based on \text{GPT-Judge}, suitable for three different scenarios: \text{EXE (Execution Evaluation)} compares the execution results between SQLs, \text{QSE (Query-SQL Evaluation)} evaluates the consistency between the query and the SQL statement by leveraging a large language model (LLM), and \text{SSE (SQL-SQL Evaluation)} uses an LLM to compare the structural and semantic consistency between two SQL statements.
Extensive offline experiments demonstrate SQLord’s effectiveness in tackling key NL2SQL challenges. Online tests also prove the scalability and robustness of SQLord in real-time query processing. SQLord has been successfully applied on the world’s largest B2B e-commerce platform, supporting essential business applications.

\section{Framework}

\subsection{Reverse Data Generation}

\begin{figure}[t]
\centering
\includegraphics[width=1.0\linewidth]{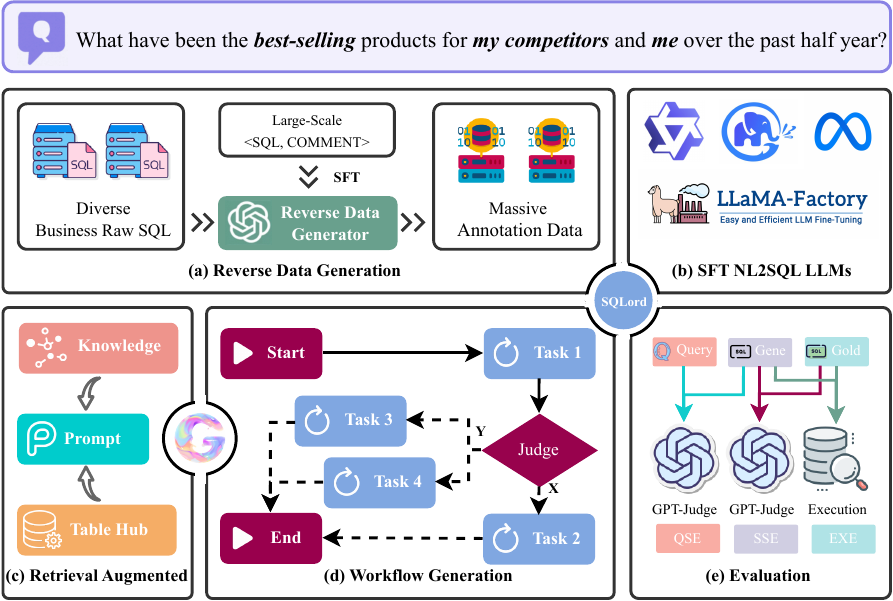}
\caption{The SQLord framework: (a) Reverse Data Generation for fine-tuning \( SQLLM \). (b) \( SQLLM \) generates domain-specific SQL queries. (c) Retrieval-Augmented Generation enables task decomposition. (d) Tasks execute based on dependencies. (e) GPT-Judge evaluates SQL quality.}
\label{fig:sqlord_pic}
\end{figure}

The scarcity of annotated \texttt{<Query, SQL>} pairs in specialized domains poses a significant challenge for NL2SQL model training. To address this limitation, we propose a reverse data generation approach as shown in Figure \ref{fig:sqlord_pic} (a) and (b). Specifically, we train a reverse generation model, referred to as \( RevLLM \), using a small set of annotated \texttt{<SQL, COMMENT>} pairs \( \mathcal{D}_{\text{train}} = \{ (s_j, c_j) \mid j=1,\dots,m \} \). These pairs are collected from routine development practices, where \(s_j\) represents the \(j\)-th SQL statement used as input and \(c_j\) represents its corresponding comment as output.

The trained \( RevLLM \) model is then used to generate large-scale pseudo-annotated \texttt{<Query, SQL>} pairs for an abundant set of raw SQL statements \( \mathcal{S} = \{ s_1, s_2, \dots, s_n \} \), also sourced from routine development. For each given SQL statement \( s_i \), \( RevLLM \) generates its corresponding natural language description \( q_i \), which serves as the query. This process can be formalized as:

\[
  q_i = \text{RevLLM}(s_i; \theta),
\]

where \( \theta \) represents the parameters of the model. The resulting dataset, \( \mathcal{D}_{\text{gen}} = \{ (q_i, s_i) \mid i=1,\dots,n \} \), serves as a pseudo-annotated corpus for downstream training.

Then, we fine-tune open-source LLMs, such as Qwen \cite{qwen2}, using the generated dataset \( \mathcal{D}_{\text{gen}} \). This fine-tuning process creates a domain-specific NL2SQL model, referred to as \( SQLLM \), optimized for the specific business scenario. The objective of fine-tuning can be formulated as minimizing the following loss function:

\[
  \mathcal{L}(\phi) = -\frac{1}{|\mathcal{D}_{\text{gen}}|} \sum_{(q, s) \in \mathcal{D}_{\text{gen}}} \sum_{t=1}^{|s|} \log p(s^t \mid q, s^{<t}; \phi),
\]

where \( \phi \) represents the parameters of \( SQLLM \), \( q \) is the query, \( s \) is the SQL statement, \( s^t \) is the \( t \)-th token in \( s \), and \( s^{<t} \) is the prefix of tokens before \( s^t \).

\subsection{Automate Workflow Generation}
The automation of complex business queries involves breaking high-level objectives into smaller, manageable sub-tasks that are executed in order of their dependencies. This can be considered as Retrieval-Augmented Generation, leveraging domain knowledge and table schema retrieval, followed by dynamic task generation to handle the sequential and parallel nature of sub-tasks.

\paragraph{Retrieval-Augmented}
As shown in Figure \ref{fig:sqlord_pic} (c). Given a user query \( q \), the first step is to retrieve relevant information from both the domain knowledge base \( \mathcal{K} \) and the schema repository \( \mathcal{T} \) (TableHub). This step provides essential context and structural information about the database, ensuring that the subsequent task decomposition aligns with both business requirements and the underlying data schema:
\[
  d_q = \text{RA}(q; \mathcal{K}, \mathcal{T}),
\]
where \( \text{RA}(\cdot) \) denotes a vector similarity-based retrieval system that gathers domain-specific knowledge and database schema details. The output \( d_q \) represents the comprehensive context.

\paragraph{Dynamic Generation}
The query \( q \) is progressively decomposed into sub-tasks \( t_1, t_2, \dots, t_k \) using a generator \( G(\cdot) \) (an LLM, such as Qwen). Each sub-task \( t_i \) depends on the previous tasks' results \( \{r_1, \dots, r_{i-1}\} \) and the information retrieved during the analysis phase. The generator \( G(\cdot) \) takes into account the context \( d_q \) and the intermediate results from prior tasks, ensuring that the decomposition adapts dynamically as the workflow progresses:
\[
  t_i = G(q, d_q, \{r_1, \dots, r_{i-1}\}).
\]
It ensures that task decomposition aligns with both the requirements of the query and the dependencies between sub-tasks.

\paragraph{SQL Generation and Execution}
For each generated sub-task \( t_i \), a corresponding SQL query \( s_i \) is produced using the trained \( SQLLM \), which incorporates domain knowledge and table schema information. The generated SQL query is then executed on the database to produce the result \( r_i \):
\[
  s_i = \text{SQLLM}(t_i, d_q; \phi), \quad r_i = \mathcal{E}(s_i),
\]
where \( \mathcal{E}(\cdot) \) represents the SQL execution engine, and \( \phi \) are the fine-tuned parameters of the model.

The tasks \( t_1, t_2, \dots, t_k \) are executed based on their dependencies, as shown in Figure \ref{fig:sqlord_pic} (d), some tasks can be executed in parallel if they are independent, while others may require sequential execution, where each subsequent task depends on the result of the previous one. This mixed execution strategy ensures efficiency and flexibility in handling the complexity of the query.

\paragraph{Result}
After executing all sub-tasks, the intermediate results \( \{r_1, r_2, \dots, r_k\} \) are aggregated to produce the final result \( r_q \), which answers the original query \( q \). The aggregation step may directly combine results or leverage an LLM to summarize and provide the final answer:
\[
  r_q = \text{Summary}(\{r_1, \dots, r_k\}),
\]

This RAG-liked approach ensures that complex business queries are resolved systematically. By dynamically generating tasks based on retrieved knowledge, and executing them with an adaptive execution strategy, the process effectively handles intricate dependencies, making it suitable for large-scale, real-world applications.

\subsection{Evaluation with GPT-Judge}

Evaluating NL2SQL models is challenging, as existing evaluation methods often struggle to effectively adapt to diverse scenarios, as well as the semantic variability (e.g., "sales >= 100" vs. "sales > 99") and field order diversity inherent in SQL. To address this, we propose a comprehensive and flexible evaluation framework with \textit{GPT-Judge} as shown in Figure \ref{fig:sqlord_pic} (e)., which supports multiple evaluation modes tailored to varying levels of information availability. This framework ensures robust assessment of generated SQL statments, even in incomplete settings.

\paragraph{Evaluation Modes}
This evaluation framework consists of three distinct evaluation modes:

1. Execution Evaluation (EXE):
When a database environment is available, the generated SQL query \( s_{\text{gen}} \) and the ground truth SQL query \( s_{\text{gold}} \) are executed against the database to compare their results. Let \( r_{\text{gen}} \) and \( r_{\text{gold}} \) represent the execution results:
\[
  \text{EXE Score} = \mathbb{I}(r_{\text{gen}} = r_{\text{gold}}),
\]
where \( \mathbb{I}(\cdot) \) is an indicator function that evaluates to 1 if the results match and 0 otherwise. This approach ensures the functional equivalence of execution results, regardless of syntactic differences.

2. Query-SQL Evaluation (QSE):
In scenarios where the ground truth SQL query \( s_{\text{gold}} \) is unavailable, GPT-Judge leverages an LLM (e.g., GPT-4) to evaluate whether the generated SQL query \( s_{\text{gen}} \) aligns with the intent of the natural language query \( q \). The model considers the context retrieved via \( \text{RA}(\cdot) \), including schema details and domain-specific knowledge, to make an informed judgment:
\[
  \text{QSE Score} = \text{LLMJudge}(q, s_{\text{gen}}, d_q),
\]
where \( \text{LLMJudge}(\cdot) \) represents the evaluation function performed by the LLM, assessing semantic consistency between \( q \) and \( s_{\text{gen}} \).

3. SQL-SQL Evaluation (SSE):
When both the generated SQL \( s_{\text{gen}} \) and the ground truth SQL \( s_{\text{gold}} \) are available, GPT-Judge evaluates their similarity directly. This includes comparing structural elements (e.g., SELECT fields, WHERE conditions) and overall semantics, which can be challenging due to variations in SQL expressions (e.g., "sales > 99" vs. "sales >= 100"):
\[
  \text{SSE Score} = \text{LLMJudge}(s_{\text{gen}}, s_{\text{gold}}).
\]

SQLord integrates three evaluation modes to provide a rigorous and adaptable assessment framework, bridging the gap between research methodologies and real-world applications while enhancing SQLord's robustness in complex scenarios.

\section{Experiments}
In this section, we conduct extensive offline evaluations on the open-source Spider \cite{spider} dataset and a manually annotated dataset from two enterprise scenarios, and validate SQLord online on the world’s largest B2B e-commerce platform across multiple real-world application scenarios.

\paragraph{Dataset} The dataset comprises 6000 \texttt{<Query, SQL>} pairs, from two online scenarios: \textit{Customs Import-Export Assistant} for trade data analysis, and \textit{Intelligent Product Selection} to assist merchants in product sourcing. These queries reflect the complexities of real-world NL2SQL tasks on the world’s largest B2B e-commerce platform. Additionally, \textit{Spider} is an open, large-scale, cross-domain text-to-SQL dataset annotated by 11 Yale students.

\paragraph{Implementation}
It was implemented with \texttt{Qwen2-7B-Instruct} (referred to simply as \texttt{Qwen}) \cite{qwen2} as the base LLM. The reverse data generation model \texttt{RevLLM} was trained on 73,589 \texttt{<SQL, COMMENT>} pairs, generating 35,948 pseudo-annotated \texttt{<Query, SQL>} pairs, which were used to fine-tune \texttt{SQLLM}. They are trained in 3 epochs with a learning rate of 1e-5 and a batch size of 1. All experiments were conducted on 8 NVIDIA H100 GPUs, with evaluation based on execution accuracy (EXE), query-SQL consistency (QSE), and SQL-SQL equivalence (SSE).

\begin{table}[h]
\centering
\caption{Offline Evaluation Results on Real-World Dataset}

\label{tab:offline_results_real}
\begin{tabular}{l|ccc}
\toprule
\textbf{Method} & \textbf{EXE (\%)} & \textbf{QSE (\%)} & \textbf{SSE (\%)} \\
\midrule
Qwen & 51.3 & 58.1 & 51.7 \\
GPT-4 & 68.5 & 75.6 & 69.2 \\
DIN-SQL (GPT) & 71.4 & 78.3 & 71.9 \\
DAIL-SQL (GPT) & 72.9 & 78.7 & 72.3 \\
SQLord (Qwen) & \textbf{86.5} & \textbf{93.2} & \textbf{85.8} \\
\bottomrule
\end{tabular}
\end{table}

\begin{table}[h]
\centering
\caption{Offline Evaluation Results on Spider Dataset}
\label{tab:offline_results_spider}
\begin{tabular}{l|ccc}
\toprule
\textbf{Method} & \textbf{EXE (\%)} & \textbf{QSE (\%)} & \textbf{SSE (\%)} \\
\midrule
Qwen & 66.7 & 72.9 & 67.1 \\
GPT-4 & 80.2 & 87.6 & 80.8 \\
DIN-SQL (GPT) & 85.3 & 91.5 & 85.9 \\
DAIL-SQL (GPT) & 86.6 & 93.4 & 86.8 \\
SQLord (Qwen) & \textbf{87.4} & \textbf{93.5} & \textbf{87.2} \\
\bottomrule
\end{tabular}
\end{table}

\subsection{Offline Evaluation}
We evaluate SQLord on two datasets: the Real-World Dataset, consisting of 6,000 annotated \texttt{<Query, SQL>} pairs from two enterprise scenarios, and the Spider. Evaluation metrics include execution accuracy (EXE), query-SQL consistency (QSE), and SQL-SQL equivalence (SSE). We compare SQLord with several baselines: Qwen; GPT-4 \cite{DBLP:journals/corr/abs-2303-08774}, using in-context learning; DIN-SQL (GPT) \cite{DBLP:conf/nips/PourrezaR23} and DAIL-SQL (GPT) \cite{gao2024text}, GPT-based frameworks; and SQLord (Qwen), with Qwen as the base LLM. Results are shown in Tables \ref{tab:offline_results_real} and \ref{tab:offline_results_spider}.

SQLord outperforms all baselines, with SQLord (Qwen) achieving the highest scores across both evaluation datasets. On the Real-World Dataset, it achieves 86.5\% EXE, 13.6 percentage points higher than DAIL-SQL (GPT), and 87.4\% on the Spider Dataset, showcasing its generalization ability. Additionally, the consistent differences in QSE and SSE scores ($\pm$7\%) confirm GPT-Judge's reliability, with a consistent gap between EXE and GPT-Judge ( $\pm$7\% for QSE and $\pm$1\% for SSE), validating its effectiveness as an execution-independent evaluation method. These results highlight SQLord’s robustness across diverse tasks.

\begin{table}[h]
\centering
\caption{Ablation Study Results on Real-World Dataset}

\label{tab:ablation_results}
\begin{tabular}{l|ccc}
\toprule
\textbf{Configuration} & \textbf{EXE (\%)} & \textbf{QSE (\%)} & \textbf{SSE (\%)} \\
\midrule
Qwen & 51.3 & 58.1 & 51.7 \\
+ Reverse Generation & 63.1 & 69.8 & 63.7 \\
+ Workflow Generation & 74.6 & 81.2 & 74.3 \\
SQLord (Qwen) & \textbf{86.5} & \textbf{93.2} & \textbf{85.8} \\
\bottomrule
\end{tabular}
\end{table}

\subsection{Ablation Studies}

To assess the contribution of SQLord's core components, we conduct ablation studies on the Real-World Dataset using Qwen as the base LLM. The configurations tested include the baseline Qwen model, followed by the addition of reverse data generation, workflow generation, and the full SQLord framework. The results are shown in Table \ref{tab:ablation_results}, demonstrating that each component significantly enhances SQLord's performance. Reverse data generation and workflow generation provide the largest improvements, with SQLord (Qwen) achieving an EXE score of 86.5\%, a QSE score of 93.2\%, and an SSE score of 85.8\%, outperforming the baseline Qwen by 35.2\%, 35.1\%, and 34.1\%, respectively.

\begin{table}[h]
\centering
\caption{Online Evaluation Results}

\label{tab:online_results}
\begin{tabular}{l|ccc}
\toprule
\textbf{Scenario} & \textbf{EXE (\%)} & \textbf{QSE (\%)} & \textbf{SSE (\%)} \\
\midrule
Customs Import-Export Assistant & +15.6 & +14.4 & +16.1 \\
Intelligent Product Selection & +16.4 & +15.9 & +16.7 \\
\bottomrule
\end{tabular}
\end{table}

\subsection{Online Evaluation}

SQLord (Qwen) was applied in two enterprise application scenarios to evaluate the accuracy of generated SQL queries in real-world environments, using execution accuracy (EXE), query-SQL consistency (QSE), and SQL-SQL equivalence (SSE) as metrics. We compare the experimental performance before and after the launch, the results are shown in Table \ref{tab:online_results}. It demonstrates SQLord’s strong performance across both scenarios with a great increase in EXE scores: +15.6\% for \textit{Customs Import-Export Assistant} and +16.4\% for \textit{Intelligent Product Selection}, highlighting its robustness and adaptability in handling complex real-world queries.

\section{Conclusion}

This paper presents SQLord, an enterprise-oriented NL2SQL framework that effectively addresses domain-specific challenges through reverse data generation, workflow decomposition, and a robust evaluation framework. Experiments on real-world datasets and successful deployments in enterprise scenarios demonstrate SQLord’s superiority over existing methods in handling complex business queries with high accuracy and efficiency. SQLord’s modular architecture ensures adaptability to evolving business needs, while its integration with large language models highlights its potential for continuous improvement and innovation in the NL2SQL domain. By bridging research and practical applications, SQLord provides a scalable foundation for advancing intelligent data systems in diverse industries.

\bibliographystyle{ACM-Reference-Format}
\bibliography{sample-base}

%%% -*-BibTeX-*-
%%% Do NOT edit. File created by BibTeX with style
%%% ACM-Reference-Format-Journals [18-Jan-2012].

\begin{thebibliography}{15}

%%% ====================================================================
%%% NOTE TO THE USER: you can override these defaults by providing
%%% customized versions of any of these macros before the \bibliography
%%% command.  Each of them MUST provide its own final punctuation,
%%% except for \shownote{}, \showDOI{}, and \showURL{}.  The latter two
%%% do not use final punctuation, in order to avoid confusing it with
%%% the Web address.
%%%
%%% To suppress output of a particular field, define its macro to expand
%%% to an empty string, or better, \unskip, like this:
%%%
%%% \newcommand{\showDOI}[1]{\unskip}   % LaTeX syntax
%%%
%%% \def \showDOI #1{\unskip}           % plain TeX syntax
%%%
%%% ====================================================================

\ifx \showCODEN    \undefined \def \showCODEN     #1{\unskip}     \fi
\ifx \showDOI      \undefined \def \showDOI       #1{#1}\fi
\ifx \showISBNx    \undefined \def \showISBNx     #1{\unskip}     \fi
\ifx \showISBNxiii \undefined \def \showISBNxiii  #1{\unskip}     \fi
\ifx \showISSN     \undefined \def \showISSN      #1{\unskip}     \fi
\ifx \showLCCN     \undefined \def \showLCCN      #1{\unskip}     \fi
\ifx \shownote     \undefined \def \shownote      #1{#1}          \fi
\ifx \showarticletitle \undefined \def \showarticletitle #1{#1}   \fi
\ifx \showURL      \undefined \def \showURL       {\relax}        \fi
% The following commands are used for tagged output and should be
% invisible to TeX
\providecommand\bibfield[2]{#2}
\providecommand\bibinfo[2]{#2}
\providecommand\natexlab[1]{#1}
\providecommand\showeprint[2][]{arXiv:#2}

\bibitem[qwe(2024)]%
        {qwen2}
 \bibinfo{year}{2024}\natexlab{}.
\newblock \showarticletitle{Qwen2 Technical Report}.
\newblock  (\bibinfo{year}{2024}).
\newblock


\bibitem[at~Yale~University(2018)]%
        {spider}
\bibfield{author}{\bibinfo{person}{LILY~Group at Yale~University}.} \bibinfo{year}{2018}\natexlab{}.
\newblock \showarticletitle{Spider 1.0, Yale Semantic Parsing and Text-to-SQL Challenge}.
\newblock  (\bibinfo{year}{2018}).
\newblock
\urldef\tempurl%
\url{https://yale-lily.github.io/spider}
\showURL{%
\tempurl}


\bibitem[Dong et~al\mbox{.}(2023)]%
        {DBLP:journals/corr/abs-2307-07306}
\bibfield{author}{\bibinfo{person}{Xuemei Dong}, \bibinfo{person}{Chao Zhang}, \bibinfo{person}{Yuhang Ge}, \bibinfo{person}{Yuren Mao}, \bibinfo{person}{Yunjun Gao}, \bibinfo{person}{Lu Chen}, \bibinfo{person}{Jinshu Lin}, {and} \bibinfo{person}{Dongfang Lou}.} \bibinfo{year}{2023}\natexlab{}.
\newblock \showarticletitle{{C3:} Zero-shot Text-to-SQL with ChatGPT}.
\newblock \bibinfo{journal}{\emph{CoRR}}  \bibinfo{volume}{abs/2307.07306} (\bibinfo{year}{2023}).
\newblock


\bibitem[Gao et~al\mbox{.}(2024)]%
        {gao2024text}
\bibfield{author}{\bibinfo{person}{Dawei Gao}, \bibinfo{person}{Haibin Wang}, \bibinfo{person}{Yaliang Li}, \bibinfo{person}{Xiuyu Sun}, \bibinfo{person}{Yichen Qian}, \bibinfo{person}{Bolin Ding}, {and} \bibinfo{person}{Jingren Zhou}.} \bibinfo{year}{2024}\natexlab{}.
\newblock \showarticletitle{Text-to-SQL Empowered by Large Language Models: A Benchmark Evaluation}.
\newblock \bibinfo{journal}{\emph{Proceedings of the VLDB Endowment}} \bibinfo{volume}{17}, \bibinfo{number}{5} (\bibinfo{year}{2024}).
\newblock


\bibitem[Hui et~al\mbox{.}(2022)]%
        {DBLP:journals/corr/abs-2203-06958}
\bibfield{author}{\bibinfo{person}{Binyuan Hui}, \bibinfo{person}{Ruiying Geng}, \bibinfo{person}{Lihan Wang}, \bibinfo{person}{Bowen Qin}, \bibinfo{person}{Yanyang Li}, \bibinfo{person}{Bowen Li}, \bibinfo{person}{Jian Sun}, {and} \bibinfo{person}{Yongbin Li}.} \bibinfo{year}{2022}\natexlab{}.
\newblock \showarticletitle{S{\({^2}\)}SQL: Injecting Syntax to Question-Schema Interaction Graph Encoder for Text-to-SQL Parsers}. In \bibinfo{booktitle}{\emph{Findings of the Association for Computational Linguistics: {ACL} 2022, Dublin, Ireland, May 22-27, 2022}}. \bibinfo{publisher}{Association for Computational Linguistics}, \bibinfo{pages}{1254--1262}.
\newblock


\bibitem[Li et~al\mbox{.}(2023b)]%
        {DBLP:conf/aaai/Li00023}
\bibfield{author}{\bibinfo{person}{Haoyang Li}, \bibinfo{person}{Jing Zhang}, \bibinfo{person}{Cuiping Li}, {and} \bibinfo{person}{Hong Chen}.} \bibinfo{year}{2023}\natexlab{b}.
\newblock \showarticletitle{Resdsql: Decoupling schema linking and skeleton parsing for text-to-sql}. In \bibinfo{booktitle}{\emph{Proceedings of the AAAI Conference on Artificial Intelligence}}, Vol.~\bibinfo{volume}{37}. \bibinfo{pages}{13067--13075}.
\newblock


\bibitem[Li et~al\mbox{.}(2023a)]%
        {DBLP:journals/corr/abs-2301-07507}
\bibfield{author}{\bibinfo{person}{Jinyang Li}, \bibinfo{person}{Binyuan Hui}, \bibinfo{person}{Reynold Cheng}, \bibinfo{person}{Bowen Qin}, \bibinfo{person}{Chenhao Ma}, \bibinfo{person}{Nan Huo}, \bibinfo{person}{Fei Huang}, \bibinfo{person}{Wenyu Du}, \bibinfo{person}{Luo Si}, {and} \bibinfo{person}{Yongbin Li}.} \bibinfo{year}{2023}\natexlab{a}.
\newblock \showarticletitle{Graphix-t5: Mixing pre-trained transformers with graph-aware layers for text-to-sql parsing}. In \bibinfo{booktitle}{\emph{Proceedings of the AAAI Conference on Artificial Intelligence}}, Vol.~\bibinfo{volume}{37}. \bibinfo{pages}{13076--13084}.
\newblock


\bibitem[Liu et~al\mbox{.}(2022)]%
        {DBLP:conf/acl-deelio/LiuSZDCC22}
\bibfield{author}{\bibinfo{person}{Jiachang Liu}, \bibinfo{person}{Dinghan Shen}, \bibinfo{person}{Yizhe Zhang}, \bibinfo{person}{Bill Dolan}, \bibinfo{person}{Lawrence Carin}, {and} \bibinfo{person}{Weizhu Chen}.} \bibinfo{year}{2022}\natexlab{}.
\newblock \showarticletitle{What Makes Good In-Context Examples for GPT-3?}. In \bibinfo{booktitle}{\emph{Proceedings of Deep Learning Inside Out: The 3rd Workshop on Knowledge Extraction and Integration for Deep Learning Architectures, DeeLIO@ACL 2022, Dublin, Ireland and Online, May 27, 2022}}. \bibinfo{publisher}{Association for Computational Linguistics}, \bibinfo{pages}{100--114}.
\newblock


\bibitem[Liu et~al\mbox{.}(2024)]%
        {DBLP:journals/corr/abs-2408-05109}
\bibfield{author}{\bibinfo{person}{Xinyu Liu}, \bibinfo{person}{Shuyu Shen}, \bibinfo{person}{Boyan Li}, \bibinfo{person}{Peixian Ma}, \bibinfo{person}{Runzhi Jiang}, \bibinfo{person}{Yuyu Luo}, \bibinfo{person}{Yuxin Zhang}, \bibinfo{person}{Ju Fan}, \bibinfo{person}{Guoliang Li}, {and} \bibinfo{person}{Nan Tang}.} \bibinfo{year}{2024}\natexlab{}.
\newblock \showarticletitle{A Survey of {NL2SQL} with Large Language Models: Where are we, and where are we going?}
\newblock \bibinfo{journal}{\emph{CoRR}}  \bibinfo{volume}{abs/2408.05109} (\bibinfo{year}{2024}).
\newblock


\bibitem[Nan et~al\mbox{.}(2023)]%
        {nan2023enhancing}
\bibfield{author}{\bibinfo{person}{Linyong Nan}, \bibinfo{person}{Yilun Zhao}, \bibinfo{person}{Weijin Zou}, \bibinfo{person}{Narutatsu Ri}, \bibinfo{person}{Jaesung Tae}, \bibinfo{person}{Ellen Zhang}, \bibinfo{person}{Arman Cohan}, {and} \bibinfo{person}{Dragomir Radev}.} \bibinfo{year}{2023}\natexlab{}.
\newblock \showarticletitle{Enhancing text-to-SQL capabilities of large language models: A study on prompt design strategies}. In \bibinfo{booktitle}{\emph{Findings of the Association for Computational Linguistics: EMNLP 2023}}. \bibinfo{pages}{14935--14956}.
\newblock


\bibitem[OpenAI(2023)]%
        {DBLP:journals/corr/abs-2303-08774}
\bibfield{author}{\bibinfo{person}{OpenAI}.} \bibinfo{year}{2023}\natexlab{}.
\newblock \showarticletitle{{GPT-4} Technical Report}.
\newblock \bibinfo{journal}{\emph{CoRR}}  \bibinfo{volume}{abs/2303.08774} (\bibinfo{year}{2023}).
\newblock
\urldef\tempurl%
\url{https://doi.org/10.48550/arXiv.2303.08774}
\showURL{%
\tempurl}


\bibitem[Pourreza and Rafiei(2023)]%
        {DBLP:conf/nips/PourrezaR23}
\bibfield{author}{\bibinfo{person}{Mohammadreza Pourreza} {and} \bibinfo{person}{Davood Rafiei}.} \bibinfo{year}{2023}\natexlab{}.
\newblock \showarticletitle{{DIN-SQL:} Decomposed In-Context Learning of Text-to-SQL with Self-Correction}. In \bibinfo{booktitle}{\emph{Advances in Neural Information Processing Systems 36: Annual Conference on Neural Information Processing Systems 2023, NeurIPS 2023, New Orleans, LA, USA, December 10 - 16, 2023}}.
\newblock


\bibitem[Rajkumar et~al\mbox{.}(2022)]%
        {DBLP:journals/corr/abs-2204-00498}
\bibfield{author}{\bibinfo{person}{Nitarshan Rajkumar}, \bibinfo{person}{Raymond Li}, {and} \bibinfo{person}{Dzmitry Bahdanau}.} \bibinfo{year}{2022}\natexlab{}.
\newblock \showarticletitle{Evaluating the Text-to-SQL Capabilities of Large Language Models}.
\newblock \bibinfo{journal}{\emph{CoRR}}  \bibinfo{volume}{abs/2204.00498} (\bibinfo{year}{2022}).
\newblock


\bibitem[Trummer(2022)]%
        {DBLP:journals/pvldb/Trummer22}
\bibfield{author}{\bibinfo{person}{Immanuel Trummer}.} \bibinfo{year}{2022}\natexlab{}.
\newblock \showarticletitle{CodexDB: Synthesizing code for query processing from natural language instructions using GPT-3 Codex}.
\newblock \bibinfo{journal}{\emph{Proceedings of the VLDB Endowment}} \bibinfo{volume}{15}, \bibinfo{number}{11}, \bibinfo{pages}{2921--2928}.
\newblock


\bibitem[Wang et~al\mbox{.}(2020)]%
        {DBLP:conf/acl/WangSLPR20}
\bibfield{author}{\bibinfo{person}{Bailin Wang}, \bibinfo{person}{Richard Shin}, \bibinfo{person}{Xiaodong Liu}, \bibinfo{person}{Oleksandr Polozov}, {and} \bibinfo{person}{Matthew Richardson}.} \bibinfo{year}{2020}\natexlab{}.
\newblock \showarticletitle{{RAT-SQL:} Relation-Aware Schema Encoding and Linking for Text-to-SQL Parsers}. In \bibinfo{booktitle}{\emph{Proceedings of the 58th Annual Meeting of the Association for Computational Linguistics, {ACL} 2020, Online, July 5-10, 2020}}. \bibinfo{publisher}{Association for Computational Linguistics}, \bibinfo{pages}{7567--7578}.
\newblock


\end{thebibliography}

\end{document}